\documentstyle[12pt,epsfig]{article}  
\def\lapproxeq{\lower .7ex\hbox{$\;\stackrel{\textstyle 
<}{\sim}\;$}} 
\def\gapproxeq{\lower .7ex\hbox{$\;\stackrel{\textstyle 
>}{\sim}\;$}} 

\def\lapproxeq{\lower .7ex\hbox{$\;\stackrel{\textstyle  
<}{\sim}\;$}}  
\def\gapproxeq{\lower .7ex\hbox{$\;\stackrel{\textstyle  
>}{\sim}\;$}}  
\def\be{\begin{equation}}                                                 
\def\ee{\end{equation}}                                                   
\def\bea{\begin{eqnarray}}                                                
\def\eea{\end{eqnarray}}                                                  

\begin{document}  
\begin{titlepage}  
\pagestyle{empty}  
 
\begin{center}  
{\large \bf Parametrisation of $F_2^\gamma$ at low $Q^2$ and of $\sigma_ 
{\gamma\gamma}$ and $\sigma_{\gamma^*\gamma}$ at high energies} \\ 
\vspace{1.1cm}  
         {\sc B.~Bade\l{}ek$^a$},  {\sc M.~Krawczyk$^b$},  
         {\sc J.~Kwieci\'nski$^c$} and  {\sc A. M.  Sta\'sto$^c$}\\  
\vspace{0.3cm}  
$^a$ {\it Department of Physics, Uppsala University, P.O.Box 530,  
751 21 Uppsala, Sweden} \\  
 
{\it and Institute of Experimental Physics, Warsaw University,  
00-681 Warsaw, Poland }\\  
  
$^b$ {\it Institute of Theoretical Physics, Warsaw University,  
00-681 Warsaw, Poland }\\

$^c$ {\it Department of Theoretical Physics,   
H.~Niewodnicza\'nski Institute of Nuclear Physics, \\  
31-342 Cracow, Poland} \\  
\end{center}  
  \vspace{2cm}  
\begin{abstract}   
A parametrisation of the real photon structure function  $F_2^{\gamma}$ 
in the low $Q^2$, low $x$ region is formulated.   
It includes both the VMD and the QCD  components, the latter 
suitably extrapolated to the low $Q^2$ region and based on  
arbitrary  parton distributions in the photon. 
The parametrisation used together with the GRV and GRS' parton densities 
describes reasonably well the existing high energy  
data   
on $F_2^{\gamma}$, $\sigma_{\gamma \gamma}$  
and the low $Q^2$ data on $\sigma_{\gamma^* \gamma}$.  
Predictions for   
$\sigma_{\gamma \gamma}$  and for $\sigma_{\gamma^* \gamma}$    
for energies which may become accessible in future linear colliders   
are also given.  
\end{abstract}  
\end{titlepage}  
\section{Introduction}  
\noindent

Electron--photon scattering, 
\begin{equation}  
e ~\gamma \rightarrow \gamma^* \gamma \rightarrow hadrons,  
\end{equation}  
studied in high energy $e^+e^-$ collisions with tagged electron  
is an analogue of the inelastic lepton--nucleon scattering.  
Here the probe -- a  virtual photon of four momentum $q$  
($q^2\;=\;-Q^2 < 0$), tests the   
target particle, the real photon of four momentum $p$ 
($p^2\;=\;0$).  
The corresponding spin averaged cross section, Fig.1, 
can be parametrized e.g. by the photon structure functions  
$F_1^{\gamma}(x,Q^2)$ and $F_2^{\gamma}(x,Q^2)$.  
The Bjorken parameter $x$  
is conventionally defined as $x \; = \; {Q^2 / (2p \cdot q})$. 
Thus the process (1) permits an insight into the inner  
structure of the real photon.\\  
 
\begin{figure}[htb] 
\centerline{\epsfig{file=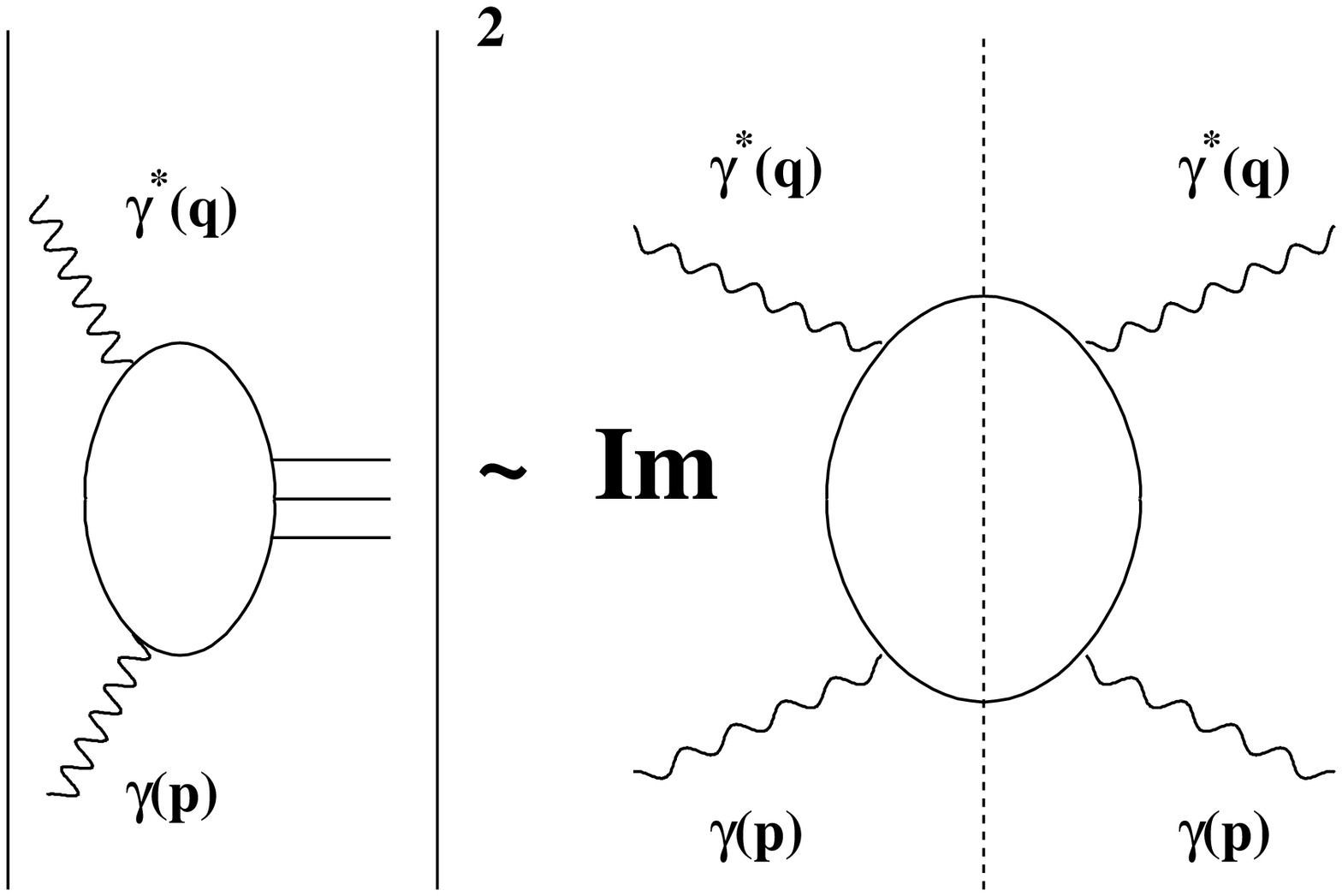,height=8cm,width=10cm}} 
\small{Figure 1: The cross section for the $\gamma^*\gamma\rightarrow  
hadrons$ scattering and its relation to the imaginary part of the 
forward $\gamma^*\gamma\rightarrow \gamma^*\gamma$ amplitude. 
} 
\label{fig:fig1} 
\end{figure} 
  
At large $Q^2$ the photon structure function is described by the  perturbative  
QCD \cite{WITTEN,ZERWAS,LAC,YELLOW,KRAWZ}. However   
in the low  $Q^2$  region, $Q^2\lapproxeq$ 1 GeV$^2$, it is expected that  
the Vector Meson Dominance (VMD) contribution \cite{VMD} is important. 
  
In this paper we   present a model of the photon structure   
function $F_2^\gamma$ which includes both the VMD contribution and the QCD  
term, suitably extrapolated to the low $Q^2$ region.  This approach   
is based on the extension of a similar representation   
of the nucleon structure function \cite{BBJK1,BBJK2,MRS} to the case of the   
photon. Possible parametrisations of the photon structure function   
which extend to the low $Q^2$ region have also been discussed in   
Ref. \cite{GLM,GALUGA,DOSCH1}.  The parametrisation proposed in Ref.\cite{GLM}  
is based upon the Quark Parton Model supplemented by the contribution  
from the hadronic structure of the photon.  The energy dependence of the  
latter has a Regge form.  The $Q^2$ dependence is parametrised in terms of the simple  
form-factors which if combined with the Regge-type energy dependence  
generate at large $Q^2$  the Bjorken scaling behaviour of the corresponding part  
of the structure function.  In Ref. \cite{GALUGA} the energy dependence of the cross-section  
is also parametrised in a Regge-like form with the $Q^2$ dependence  
specified by the suitable form factors which contain  terms  
corresponding to the VMD contribution.  The parametrisation discussed in   
\cite{DOSCH1} is based upon a model corresponding to the interaction of colour  
dipoles, i.e. the $q \bar q$ pairs   
which the photon(s) fluctuate into.   In our approach    
the contribution  
coming from the light vector mesons within the Vector Meson Dominance model is  
similar to that used by other authors (see eg. \cite{GALUGA}) although  
the details concerning estimate of the relevant total cross-sections are slightly  
different.  The novel  
feature of our model is the treatment of the contribution coming from high  
masses of the hadronic states which couple to the virtual photons. In our 
scheme this contribution is directly related to  the photon structure  
function in  the large  
$Q^2$ region.   
The low and high mass hadronic states are separated at $Q_0$, a  
 parameter whose value was taken identical with that for the $F_2^{p}$.   
     
Our framework permits to describe the $\gamma\gamma$ and $\gamma^*\gamma$ 
total cross sections as functions of energy. The energy 
dependence of $\sigma_{\gamma\gamma}$ is also described by other 
models, \cite{GALUGA,DOSCH1,FS,SSGG,DOSCH2,CGP,Donnachie,Gotsman}.   
Most of them incorporate the Regge-like parametrisation of the total  
$\gamma \gamma$  cross-sections; some  provide a detailed  
insight into the structure of final states  
and a decomposition of the $\gamma \gamma $ total cross-section into terms  
corresponding to the appropriate subdivision of photon  
interactions and event classes 
\cite{GALUGA,SSGG}.  Possibility that part of the $\gamma \gamma$ cross-section is driven by the production  
of minijets has been discussed in \cite{FS,CGP}.    
Certain approaches analyse the behaviour of the cross-sections 
on the virtualities  
of both interacting photons \cite{GALUGA,DOSCH1,DOSCH2,Donnachie,Gotsman}. 
  
The content of our paper is as follows:  
In the next section we recall the QCD  
description of the photon structure functions and  
in section 3  we briefly describe the Vector Meson Dominance model in the process $\gamma^* \gamma \rightarrow hadrons$. 
In section 4 we present a parametrisation of the photon  
structure function, as well as  the total  
cross section for the interaction of two real photons and of the  
virtual and real photon.  
In section 5 we compare our theoretical predictions with the experimental  
data on the $F_2^{\gamma}$ and on the total cross sections 
 $\sigma_{\gamma \gamma}$ and $\sigma_{\gamma^* \gamma}$.  
We also give predictions for   
 $\sigma_{\gamma \gamma}$ in the very high energy range which  
can become accessible in future linear colliders.  
Finally in section 6 we give the summary of our results.

\section{Partonic content of the photon}  
In the large $Q^2$, i.e. in the deep inelastic limit the virtual photon   
probes the quark (antiquark) structure of the  
(real) photon in analogy to the deep inelastic lepton--hadron scattering, 
 Fig.2.  
\begin{figure}[htb]  
\centerline{\epsfig{file=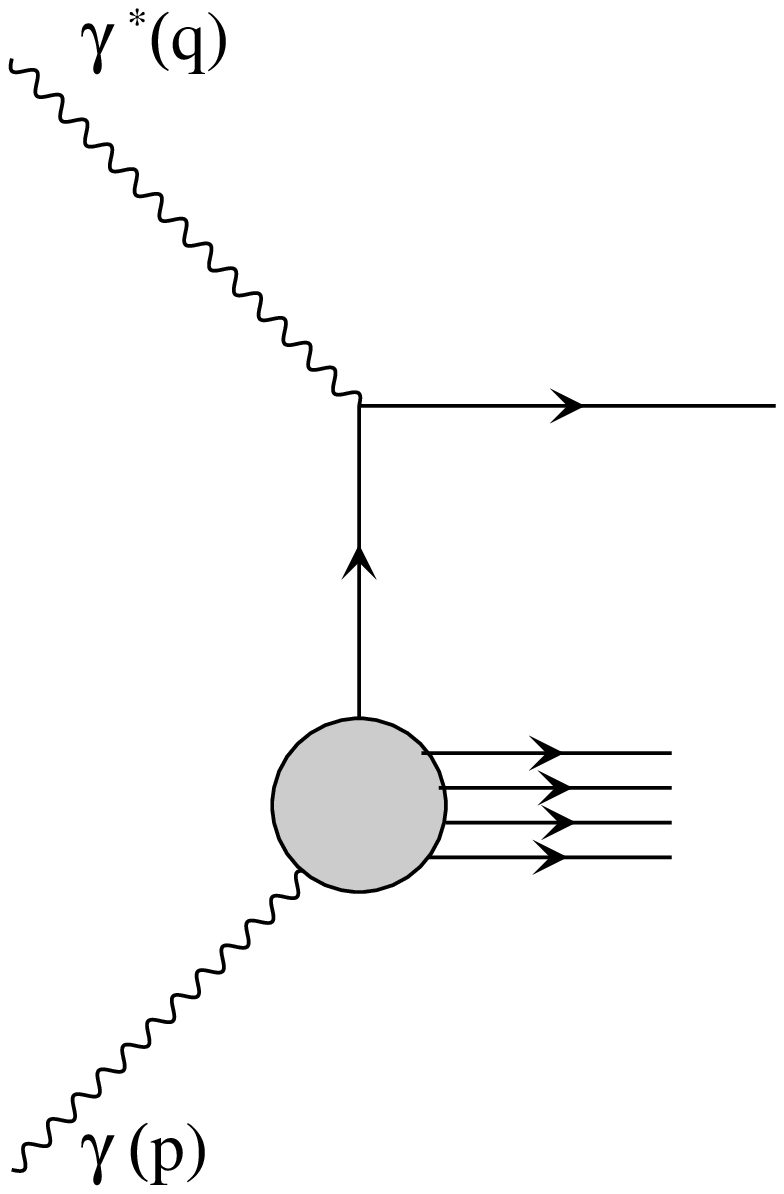,height=8cm,width=6cm}}  
\small{Figure 2: $\gamma^* \gamma$ scattering as a mechanism 
for probing partonic structure of the real photon 
} 
\label{fig:fig2}  
\end{figure}   
The corresponding photon structure function  
$F_2^{\gamma}(x,Q^2)$ may thus be   related  
to quark and antiquark distributions $q_i^\gamma (x,Q^2)$,  
$\overline{q}_i^\gamma (x,Q^2)$  
in the photon:  
\be  
F_2^{\gamma}(x,Q^2) 
= \; x \, \sum_i \; e_i^2 \; [ \; q_i^\gamma (x,Q^2)  
\; + \; \bar{q}_i^\gamma (x,Q^2)],  
\label{eq:a1}  
\ee  
where $e_i$ denote the charges of quarks and antiquarks and the sum  
is over all active quark flavours.   To be precise equation (\ref{eq:a1})  
holds   
in leading logarithmic approximation of perturbative QCD. It acquires  
higher order  
corrections in next-to-leading approximation and beyond \cite{GRV,GRS}.

A special feature of the quark structure of the photon  
with respect to the proton is a possibility of   
a direct $q \bar{q}$ production through the process  
$\gamma^* \gamma \rightarrow q \bar{q}$, see Fig.3, 
leading to the parton model predictions for the $ q_i^\gamma (x,Q^2)$ and  
 ${\bar q}_i^\gamma (x,Q^2)$. A contribution   
of this process introduces an inhomogeneous term into   
the equation describing the QCD evolution of the   
 quark (and antiquark) distributions in the photon.  
\begin{figure}[htb]  
\centerline{\epsfig{file=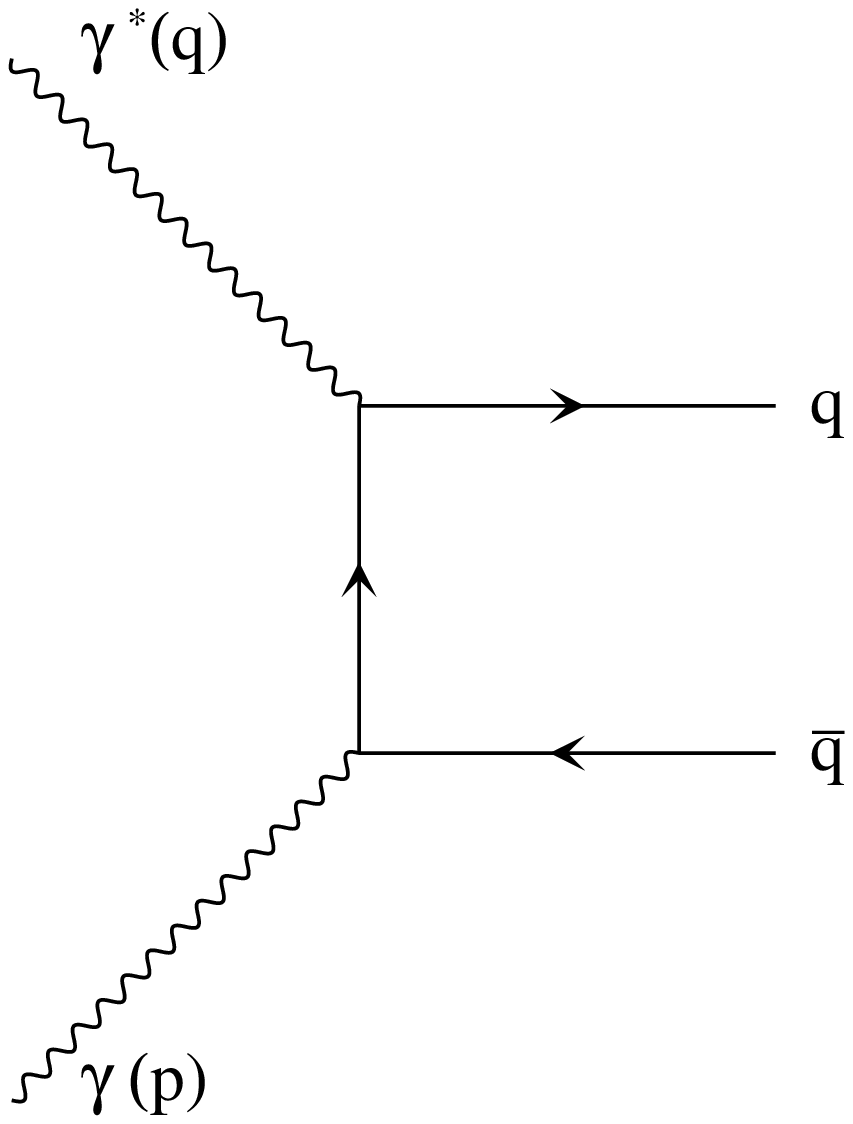,height=7cm,width=5cm}}  
\small{Figure 3: Diagrammatic representation of the direct 
$\gamma^* \gamma \rightarrow q\bar{q}$ process.  
}    
\label{fig:fig3}  
\end{figure}   
 The process $\gamma^* \gamma \rightarrow q \bar{q}$   
 (modified by  the QCD evolution)  
with the pointlike quark coupling both to real and virtual photons  
dominates in the large $Q^2$ limit making the photon structure functions exactly calculable  
in this limit \cite{WITTEN,ZERWAS}.   
Striking features of these functions are: 
 \begin{itemize} 
\item  $~~~~~~~F_{1,2}^{\gamma}$ rise with increasing $x$  
at large $x$; 
\item  $~~~~~~~F_{1,2}^{\gamma}$ show the scaling violation,  
$F_{1,2}^{\gamma} \sim \ln Q^2$.  
\end{itemize} 
At low values of $x$  
the dominant role in the photon structure functions 
is played by gluons. 
The situation here is similar to that of  
the hadronic structure functions which  
exhibit a very strong increase with decreasing $x$, see e.g. \cite{LOWX}.  
 Those effects are still rather weak in the  
kinematical region of $F_2^\gamma$ probed by present experiments  
but they will be very important in the regime  
accessible in the future linear $e^+ e^-$   
($e \gamma$, $\gamma\gamma$) colliders \cite{PC}.  
  
Besides the direct, point-like coupling to quarks, the target  
photon can fluctuate into vector mesons and other hadronic states 
which can also have their partonic structure.  
The latter cannot be calculated perturbatively  
and thus has to be parametrised phenomenologically \cite{SSGGST}.

Finally it should be pointed out that the charm quark playing 
 the dominant role in  the heavy quark contributions to  
$F_2^{\gamma}(x,Q^2)$,    
  is often described just  
by the lowest order Bethe-Heitler cross-section  
for the process $\gamma^* \gamma \rightarrow c \bar c$ and the   
additional contribution generated by the radiation $g \rightarrow c \bar c$  
\cite{GRS,SSGGST}.    
\section 
{Dispersive relation for $\gamma^* \gamma$ 
 scattering. $F_2^{VMD}$ and $F_2^{partons}$ contributions to  $F_2^\gamma$} 
%
The QCD describes the photon structure functions in the large $Q^2$  
region. In the low $Q^2$  region however one expects  
that the VMD mechanism is important.  
By the Vector Meson Dominance mechanism in this case we understand  
the model in which the virtual  
photon of virtuality $Q^2$ fluctuates into vector mesons  
which next undergo  interaction  
with the (real) photon of virtuality $p^2 \simeq 0 $, see Fig.4. 
 In order to be able to describe the photon  
structure function for arbitrary values of $Q^2$   
  it would  be very useful to have a unifed scheme  
which  contains both the VMD and the QCD contributions, the latter  
suitably extended to the region of low values of $Q^2$. 
\vspace*{-2cm} 
\begin{figure}[htb]  
\centerline{\epsfig{file=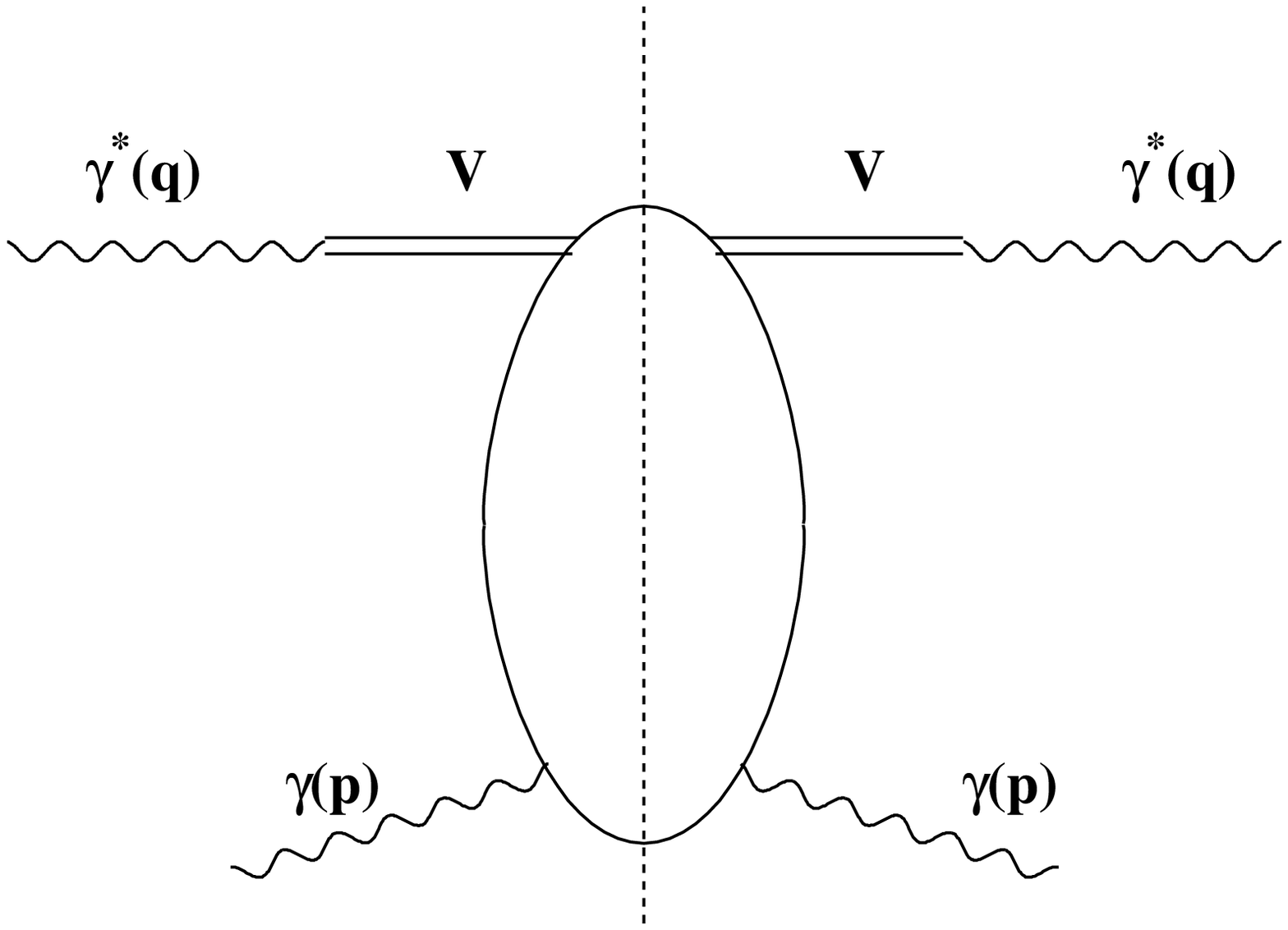,height=10cm,width=10cm}}  
\small{Figure 4: Diagrammatic repesentation of the Vector Meson Dominance model 
in the $\gamma^* \gamma$ scattering.  
}    
\label{fig:fig4}  
\end{figure}  
This may be achieved by utilising the dispersive representation  
in $Q^2$ of the structure function.  
To this aim let us notice that  
the $\gamma^* \gamma$ collision can be viewed as the interaction  
of a real photon target with a 
photon with virtuality $Q^2$ which fluctuates onto general   
hadronic state, cf. Fig.5.  
We consider the virtual photon first fluctuating  
onto the $q\bar{q}$ state and then interacting with the real photon.  
Like in the $\gamma^* p $ scattering one can write the dispersion  
relation for the $\gamma^* \gamma$ scattering as follows \cite{VMD}:  
$$                                                  
F_2^\gamma (W^2, Q^2) =  
$$  
\be  
{Q^2 \over 4 \pi^2 \alpha} \; \sum_q \: \int \: \frac{d M^2}{M^2 + Q^2} \: \int                                                     
\: \frac{d M^{\prime 2}}{M^{\prime 2} + Q^2} \; \rho (M^2,                
M^{\prime 2}) \: \frac{1}{W^2} \; {\rm Im} \; A_{(q \overline{q}) -       
\gamma}                                                     
(W^2, M^2, M^{\prime 2})       
\label{eq:aa1}                                                            
\ee                                             
where $M$ and $M^\prime$ are the invariant masses of the incoming and outgoing                                                  
$q\bar{q}$ pair.  In eq. (\ref{eq:aa1}), $\rho (M^2,M'^2)$ is the density matrix  of the $q\bar{q}$ states   
and ${\rm Im} A_{(q\bar{q})-\gamma}$ is an imaginary part of the corresponding   
forward scattering amplitude.  
  
The above formula can be rewritten in a form  
of a single dispersion relation as follows,       
\be                                                     
\label{a2} 
F_2^\gamma (W^2, Q^2) \; = \; Q^2 \: \int_0^\infty \: \frac{d Q'^2}{(Q'^2 +                                                     
Q^2)^2} \; \Phi (W^2, Q'^2) \:                                            
\ee                                                     
where  the spectral function 
\bea  
\Phi(W^2,Q'^2) & = & {1 \over 4 \pi^2 \alpha} \: \int_0^1 \: d \lambda \int dM^2 \int dM'^2  
\: \delta(Q'^2-\lambda M'^2-(1-\lambda) M^2) \nonumber \\  
& & \rho(M^2,M'^2) \: {1 \over W^2} \: {\rm Im} A_{(q\bar{q})-\gamma}(W^2,M^2,M'^2).  
\label{sindisp}  
\eea

The centre-of-mass energy squared $W^2=(p+q)^2$ is related  
to the Bjorken parameter $x$ in the following way:  
\be  
W^2 \; = \; Q^2 ({1 \over x} - 1).  
\ee  
One can now separate regions of low- and high values of $Q'^2$ 
in the integral (\ref{a2}), by noticing that this integral corresponds 
to the (Generalised) Vector Meson Dominance representation of the 
$F_2^\gamma$. 
 
\begin{figure}[htb]  
\centerline{\epsfig{file=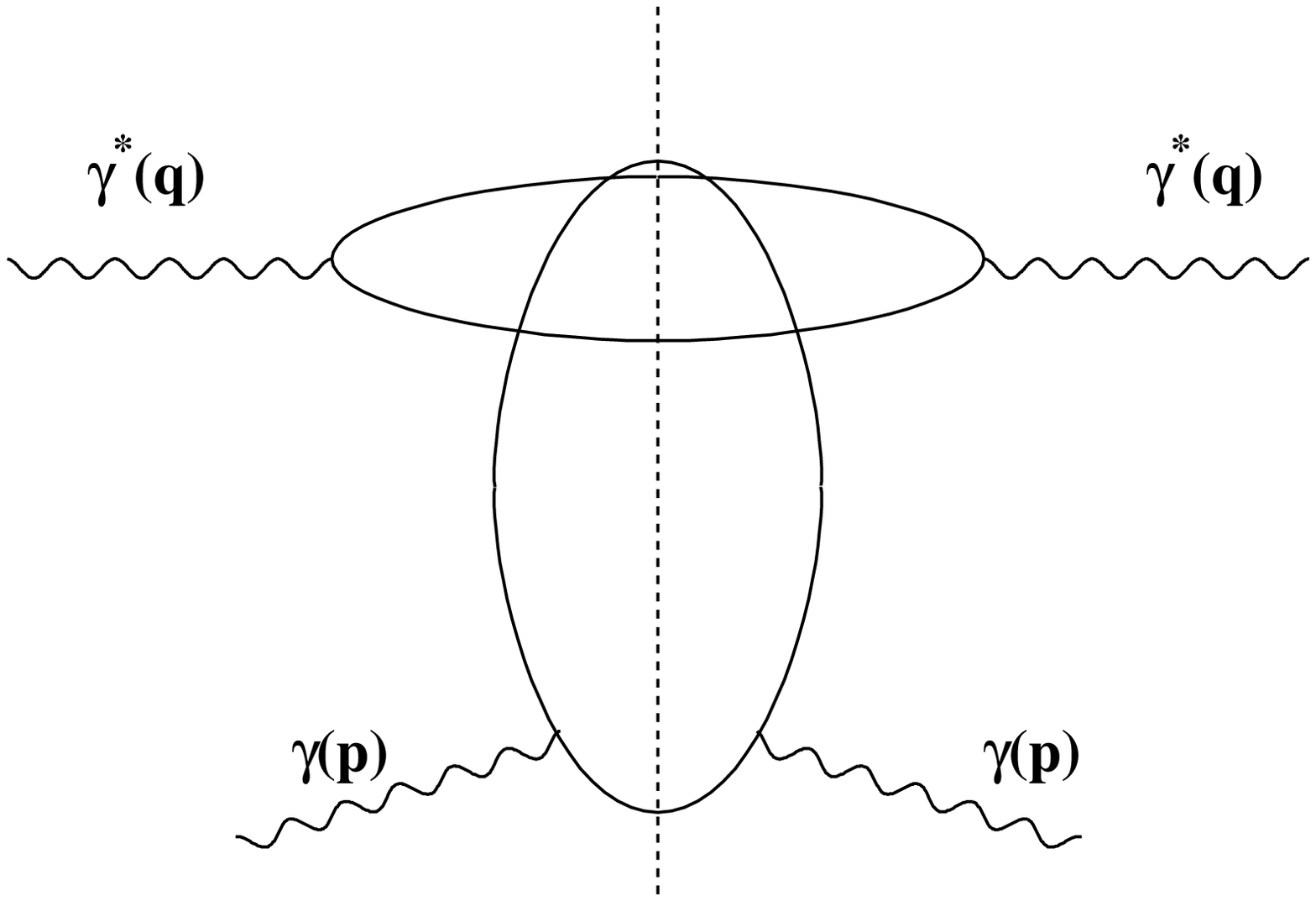,height=10cm,width=10cm}}  
\small{Figure 5: Diagrammatic representation of the dispersion relation. 
}    
\label{fig:fig5}  
\end{figure}   
For low values of $Q'^2$, 
 $Q'^2 < Q_0^2$, one uses the Vector Meson Dominance model.  
In this approach one assumes that the virtual photon  forms  
a vector meson rather than  
a pair of well separated $q~{\rm {and}}~ \bar{q}$.   
The integrand in the integral on the right hand side  
of equation (\ref{sindisp}) defining the spectral function $\Phi(W^2,Q^{\prime 2})$  
is then given by the following formula:  
$$  
\rho(M^2,M'^2) \: {1 \over W^2} \: {\rm Im} A_{(q\bar{q})-\gamma}(W^2,M^2,M'^2)  
$$ 
\begin{equation} 
=\pi \alpha \sum_{V}{M_V^4\over \gamma_V^2}\sigma_{V \gamma}  \; (W^2) 
\delta(M^2 - M_V^2) \delta(M'^2 - M_V^2), 
\label{spfvm} 
\end{equation} 
where $M_V$ is the mass of the vector meson $V$ and $\sigma_{V \gamma}(W^2)$ denotes the   
$V \gamma $ total cross section.          
The couplings  $\gamma_V^2$ can be estimated from the leptonic widths of the vector mesons,  
\begin{equation}  
 {\gamma_V^2 \over \pi}= {\alpha^2 M_V\over 3 \Gamma^V_{e^+e^-}} \; . 
\label{gammav}  
\end{equation}  
In equation (\ref{spfvm}) we have included only diagonal transitions between the vector mesons  
having the same masses.  The corresponding spectral function  
$\Phi^{VMD}(W^2,Q^{\prime 2})$ thus reads:  
\begin{equation} 
\Phi^{VMD}(W^2,Q^{\prime 2})= 
\sum_{V}{M_V^4\over 4 \pi \gamma_V^2}\sigma_{V \gamma}  \; (W^2)\delta(Q^{\prime 2} -  
M_V^2). 
\label{fivdm} 
\end{equation} 
The resulting VMD  contribution to the  
photon structure function $F_2^{\gamma}$ 
, $F_2^{\rm VMD}$, is given by the following equation:  
\be                                                     
\label{eq:a2}                                                     
F_2^{\rm VMD} (x,Q^2) \; = \; \frac{Q^2}{4 \pi} \; \sum_V \:            
\frac{M_V^4 \: \sigma_{V \gamma}  \; (W^2)}{\gamma_V^2 (Q^2 + M_V^2)^2} \; .                                                     
\ee                                             
 
In eq.(\ref{eq:a2})  we only consider mesons with masses $M_V^2< Q_0^2$.  
The contribution coming from the  region of high values of $Q'^2$ ($Q'^2 > Q_0^2$)  
can be related to the photon structure function from the large $Q^2$  domain.  
It defines the partonic contribution $F_2^{\rm partons}$ to  
the structure function $F_2^\gamma$, extended to arbitrary low values of 
$Q^2$. For convenience, we adopt  
the approximation used in Ref. \cite{BBJK2} which gives:  
\be  
F_2^{\rm partons}(x,Q^2) \; = \; {Q^2 \over Q^2 + Q_0^2} \; F_2^{\rm QCD}(\bar{x},Q^2 + Q_0^2)   
\label{f2partons}  
\ee  
where  
\be  
\label{xbar}  
\bar{x} \; = \; {Q^2 + Q_0^2 \over W^2 + Q^2 + Q_0^2} \; .  
\ee  
  
The structure function $F_2^{\rm QCD}$ is taken from the QCD  
analysis, valid in the large $Q^2$ region, i.e. it is calculated  
from the existing parametrisations of the parton distributions.  
Modifications of the QCD contribution:  
replacement of the parameter $x$ by $\bar x$ defined in equation   
(\ref{xbar}), shift of the scale $Q^2 \rightarrow Q^2 + Q^2_0$ and the   
factor $Q^2/(Q^2+Q_0^2)$ instead of 1, introduce   
power corrections which vanish as $1/Q^2$ and are   
negligible at large $Q^2$.  
The magnitude of $Q_0^2$  is set to $1.2 ~\rm GeV^2$ as in the 
case of the proton \cite{BBJK1,BBJK2}.  The $F_2^{\rm partons}$ thus defines 
a contribution to $F_2^\gamma$ at arbitrary low values of $Q^2$. 
 
An elaborated treatment of the partonic contribution to the proton structure function 
has been developed in Ref. \cite{MRS},  
where long and short distance components have been carefully separated.  
According to that paper the low mass region is dominated by the   
$q\bar{q}$ pairs with large transverse sizes  
in the impact parameter space (thus corresponding to the VMD)  
whereas the QCD part is  
dominated by  pairs of small transverse size.

\section{Parametrisation of the photon structure function  
and of the total photon--photon interaction cross sections }  
\noindent  
Our representation of the  photon structure function $F_2^\gamma (x,Q^2)$ is based  
on the following decomposition:    
\begin{equation}    
  F_2^\gamma (x,Q^2) = F_2^{\rm VMD}(x,Q^2) + F_2^{\rm partons}(x,Q^2)  
\label{dec}  
\end{equation}  
where $F_2^{\rm VMD}$ and $F_2^{\rm partons}(x,Q^2)$  
are defined by eq. (\ref{eq:a2}) and (\ref{f2partons}).  
A total $\gamma^* \gamma$ cross-section in the high energy limit is 
given by 
\begin{equation}  
\sigma_{\gamma^* \gamma}(W,Q^2)=  {4 \pi^2 \alpha \over Q^2} F_2^\gamma (x,Q^2),  
\label{sigma}  
\end{equation}  
with $x={Q^2 /( Q^2 + W^2)}$.  
The $Q^2=0$ (for fixed $W$) limit of eq. (\ref{sigma}) gives the total 
cross-section $\sigma_{\gamma \gamma}(W^2)$ corresponding to the interaction   
of two real photons.    
>From (\ref{dec}),  
(\ref{eq:a2}) and (\ref{f2partons})  we obtain the following  
expression for this cross-section at high energy:       
  
\begin{equation}  
  \sigma_{\gamma \gamma}(W) = \alpha \pi \sum_{V=\rho,\omega,\phi} 
  {\sigma_{V \gamma}(W^2)\over \gamma_V^2} + {4 \pi^2 \alpha \over Q_0^2}   
  F_2^{\rm QCD}(Q_0^2/W^2,Q_0^2) \; .  
\label{sgamgam}  
\end{equation}  
  
At large $Q^2$ the structure function given by eq. (\ref{dec})   
becomes equal   
to the QCD  contribution $F_2^{\rm QCD}(x,Q^2)$.    
The VMD component gives the power correction term which vanishes as $1/Q^2$   
for large $Q^2$.    
It should be noted that the VMD part contains only finite number of vector  
mesons with their masses smaller than $Q_0^2$.  
 
In the quantitive analysis of the photon structure function and of the  
total cross sections  we have taken the   structure function  
  $F_2^{\rm QCD}$ from the LO analyses presented in Ref. \cite{GRV} (GRV) 
 and \cite{GRS} (GRS'),with a number 
of active flavours equal four. The latter parton parametrization is based on updated  
data analysis and holds for both virtual and real photons.  
    
The VMD part was estimated using the following assumptions:  
\begin{enumerate} 
\item  
Numerical values of the couplings $\gamma_V^2$ are   
the same as those used in Ref. \cite{BBJK1}. They   
were estimated from the relation (\ref{gammav})   
which gives the following values:   
\begin{equation}  
{\gamma_{\rho}^2\over \pi}=1.98, ~~~~~~~ {\gamma_{\omega}^2\over \pi} = 21.07,   
 ~~~~~~~ {\gamma_{\phi}^2\over \pi} = 13.83.  
 \label{numbers}  
 \end{equation}  
   
\item Cross-sections     
  $\sigma_{V \gamma}$ are represented as sums of  the Pomeron and Reggeon   
  contributions:   
\begin{equation}  
 \sigma_{V \gamma}(W^2) = P_{V \gamma}(W^2) + R_{V \gamma}(W^2)   
\label{rpgv}  
\end{equation}   
where  
\begin{equation}  
 P_{V \gamma}(W^2) = a^P_{V \gamma}\left({W^2\over W_0^2} \right)^{\lambda_P}  
\label{pgv}  
\end{equation}  
\begin{equation}  
 R_{V \gamma}(W^2) = a^R_{V \gamma}\left({W^2\over W_0^2} \right)^{\lambda_R}   
\label{rgv}  
\end{equation}  
with  
\begin{equation}   
\lambda_R=-0.4525, ~~~~~~~~~~~ \lambda_P= 0.0808  
\label{lambda}  
\end{equation}   
and $W_0^2 = $1 GeV$^2$ \cite{DL}.  
\item Pomeron couplings $a^P_{V \gamma}$ are related to the corresponding   
couplings $a^P_{\gamma p}$ controlling the Pomeron contributions to the  
total $\gamma p$   
cross-sections. We assume the additive quark model and reduce the total   
cross-sections for the interaction of strange quarks by a factor 2.    
This gives:   
$$   
 a^P_{\rho \gamma}= a^P_{\omega \gamma}={2\over 3}a^P_{\gamma p}~,  
$$  
  
\begin{equation}  
a^P_{\phi \gamma}={1\over 2}a^P_{\rho \gamma}~.  
\label{pvm}  
\end{equation}  
\item Reggeon couplings $a^R_{V \gamma}$ are  estimated assuming  
the additive quark  
 model and the duality (i.e. a dominance of planar quark diagrams).  
We also assume  that the quark couplings to a photon are   
 proportional to the quark charge with the flavour independent proportionality   
 factor.  This gives:    
 $$  
 a^R_{\rho \gamma}=a^R_{\omega \gamma}={5\over 9}a^R_{\gamma p}~,  
 $$  
   
 \begin{equation}  
  a^R_{\phi \gamma}=0~.  
  \label{rvm}  
  \end{equation}  
\item Couplings $a^P_{\gamma p}$ and  $a^R_{\gamma p}$ are taken from   
the fit discussed in Ref. \cite{DL} which gave:   
\begin{equation}     
a^R_{\gamma p} = 0.129 {\rm \; mb},~~~~~~~~~ a^P_{\gamma p} = 0.0677 {\rm \; mb}   
\label{apr}  
\end{equation}

\end{enumerate}  

Since we are  using the Regge description of    
total cross sections $\sigma_{V \gamma}(W)$ our approach can only work  
 for large values of $W$,  
$W^2 \gapproxeq\; 2 ~{\rm GeV}^2$, away from the resonance region.  
\section{Numerical results}  
In this section we compare our results for the real photon structure  
function $F_2^\gamma(x,Q^2)$ and the two-photon cross sections,  
$\sigma_{\gamma\gamma}(W)$ and $\sigma_{\gamma^*\gamma}(W)$   
with corresponding measurements.  
In some cases our predictions have been extended to the region  
$ W^2 < 2~ \rm GeV^2$  where the  model may not be applicable.  
Theoretical curves were obtained  
using two different parametrisations for the structure function $F_2^{\rm QCD}$, 
 GRV \cite{GRV} 
and GRS' \cite{GRS}. 
 
In Fig.6 
 we show predictions for the photon structure function based  
on equation (\ref{dec})  
plotted as the function of $x$ for different values of $Q^2$ in the region  
of small $Q^2$. In Fig.7 
 the $Q^2$ dependence of the photon structure function  
for different values of $x$ is presented 
\footnote{  
In both figures the curves are plotted only for $W>2m_{\pi}$ which  
corresponds to the threshold energy 
in the reaction $\gamma \gamma \rightarrow hadrons$.}. 
We confront our theoretical results  
with existing experimental data \cite{PLUTO1,TPC1,L31,OPAL1,OPAL2}.  
Measurements of $F_2^{\gamma}$ are 
scarce, especially for low values of $Q^2$. 
However it can be seen that  
our prediction reproduces well the data independently of the  
parametrisation (GRV or GRS') of $F_2^{\rm QCD}$ used in the model.  
Irregular behaviour of the dashed lines observed in Figs 6 and 7 at high values  
of $x$ is connected with the teatment of the charm contribution to  
$F_2^\gamma$ in the GRS' approach. 
 
In Fig.8 we compare our predictions  with the data on   
$\sigma_{\gamma \gamma} (W)$.  
\begin{figure}[htb]  
\centerline{\epsfig{file=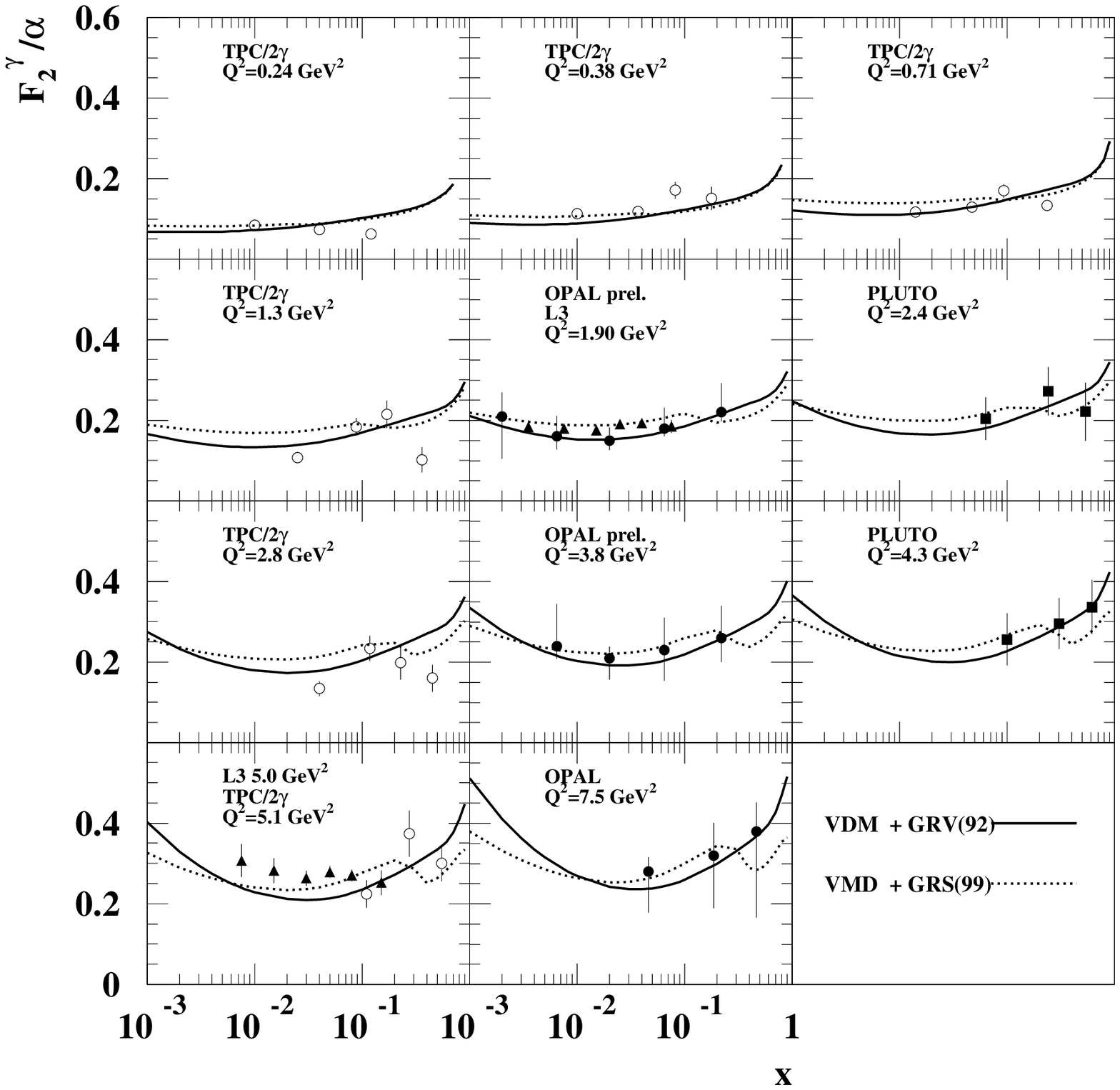,height=14cm,width=14cm}}  
\small{Figure 6: Comparison of our  predictions for $F_2^{\gamma}/\alpha$  
as a function of $x$ for different values of $Q^2$ in the low $Q^2$ region,   
with  experimental results \cite{PLUTO1,TPC1,L31,OPAL1,OPAL2}.  
The curves correspond to different LO parametrisations  
of $F_2^{\rm QCD}/\alpha$: GRV \cite{GRV} (solid line) and GRS' \cite{GRS}  
(dotted line). Error bars correspond to statistical and systematic errors added 
in quadrature.  
}    
\label{fig:fig6}  
\end{figure}   
Theoretical curves were obtained from equation (\ref{sgamgam}).  
%
We show experimental points corresponding to  
the low energy region ($W\lapproxeq $ 10 GeV) \cite{PLUTO,TPC,MD1}  and    
the recent  
preliminary high energy data   
obtained by the L3, OPAL and DELPHI collaborations at LEP \cite{L3,OPAL,DELPHI}. 
The representation (\ref{sgamgam}) for   
the total $\gamma \gamma$ cross-section describes the data reasonably well.    
The result of the calculation based on GRS' parametrisation of $F_2^{\rm QCD}$  
 is slightly higher and has a shallower minimum  
as compared to that based on GRV parametrisation.  
Calculations using the latter give  a good description  
of the shape of the energy dependence of the cross section although  
the overall normalisation seems to be about $15\%$ too large.   
 \begin{figure}[htb]  
\centerline{\epsfig{file=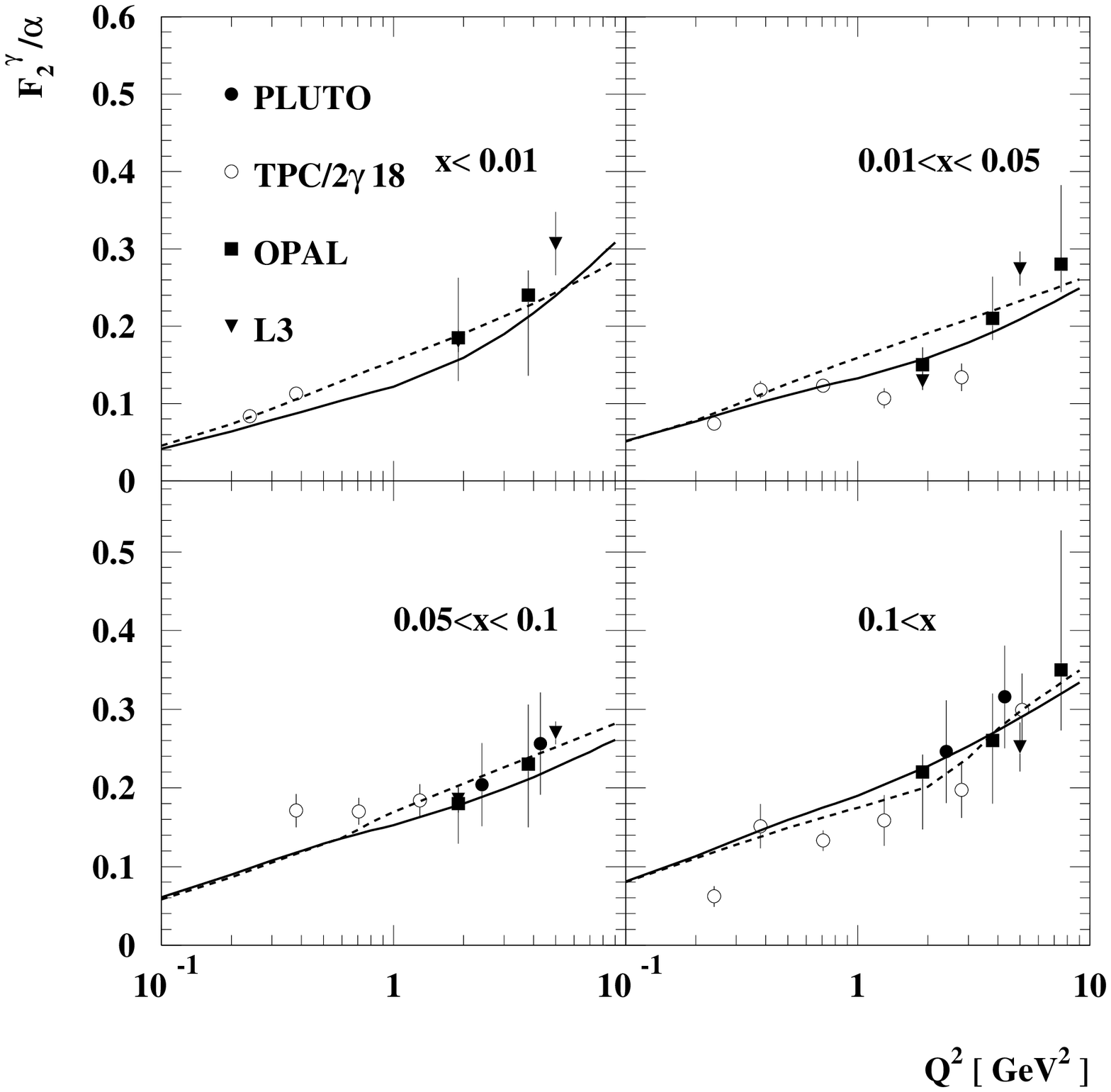,height=12cm,width=12cm}}  
\small{Figure 7: Comparison of  our predictions for $F_2^{\gamma}/\alpha$, 
as a function of $Q^2$ in the low $Q^2$ region, for different intervals of $x$, 
with experimental results \cite{PLUTO1,TPC1,L31,OPAL1,OPAL2}.  
The curves correspond to different LO parametrisation  
of $F_2^{\rm QCD}/\alpha$: GRV \cite{GRV} (solid line) and GRS' \cite{GRS}  
(dashed line). Curves are plotted for the following values  
of $x$ (clockwise): 0.0075, 0.0317, 0.0790 and 0.2617. 
Error bars correspond to statistical and systematic errors added 
in quadrature.  
 
}    
\label{fig:fig7}  
\end{figure}   
It should   
be stressed that our prediction is essentially parameter free.   
The magnitude of the cross-section is  dominated by the   
VMD component yet the partonic part is also non-negligible.   
In particular the latter   
term is responsible for generating a steeper increase of the total   
cross-section with increasing $W$ than that embodied in the VMD part which   
is described by the soft Pomeron contribution.  The decrease of the   
total cross-section  with increasing energy   
in the low $W$ region is controlled by the Reggeon component of the   
VMD part (see eqs. (\ref{rpgv}), (\ref{rgv}) and (\ref{lambda}))  
 and by the valence part of the partonic contribution.   
  
\begin{figure}[htb]  
\centerline{\epsfig{file=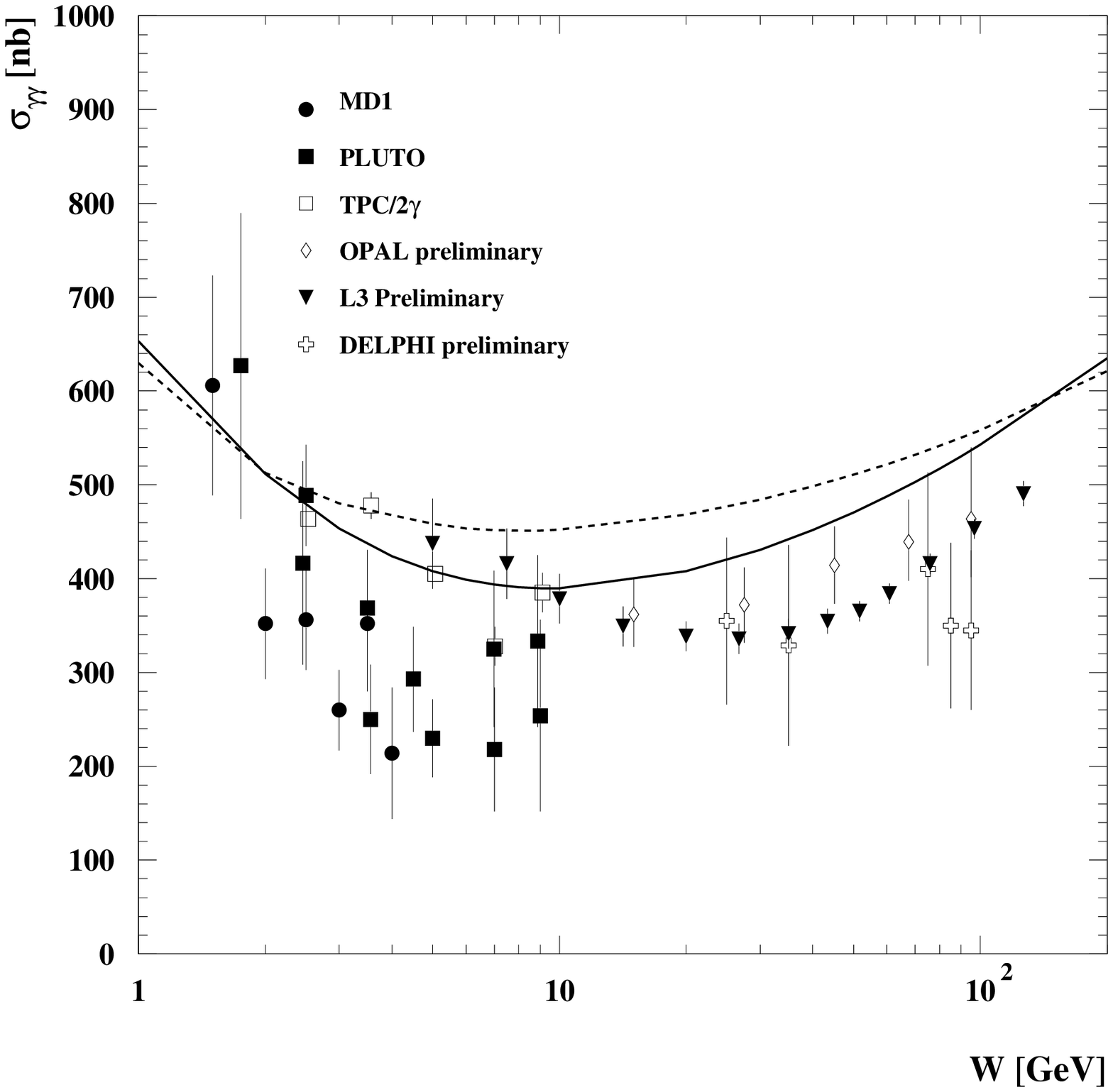,height=12cm,width=12cm}}  
\small{Figure 8: Comparison of our predictions for $\sigma_{\gamma \gamma}(W)$  
based on equation (\ref{sgamgam}) with experimental results  
 \cite{PLUTO,TPC,MD1,L3,OPAL,DELPHI}. The curves correspond to different LO  
parametrisations 
of $F_2^{\rm QCD}$: GRV \cite{GRV} (solid line) and GRS' \cite{GRS}  
(dashed line). Error bars correspond to statistical and systematic errors added 
in quadrature.  
} 
\label{fig:fig8}  
\end{figure}   
  
In Fig.9 we show predictions for the total $\gamma \gamma$ cross-section  
 as a function of the total centre-of-mass energy $W$ in the wide energy range   
including the energies that might be accessible in the future   
 linear colliders.    
In this figure we also show a decomposition of $\sigma_{\gamma \gamma}  
(W^2)$  into its VMD and partonic components (only GRV parametrisation 
was used in this analysis). At very   
high energies these two terms   
exhibit different energy dependence.  The VMD part is described   
by the soft Pomeron contribution which gives  the $W^{2 \lambda}$ behaviour   
with $\lambda$ = 0.0808, eq. (\ref{lambda}). The partonic component increases   
faster with energy since  its energy   
dependence reflects the increase of $F_2^{\rm QCD}(\bar x,Q_0^2)$ with decreasing   
$\bar x$ generated by the QCD evolution \cite{GRV,GRS}.  
\begin{figure}[htb]  
\centerline{\epsfig{file=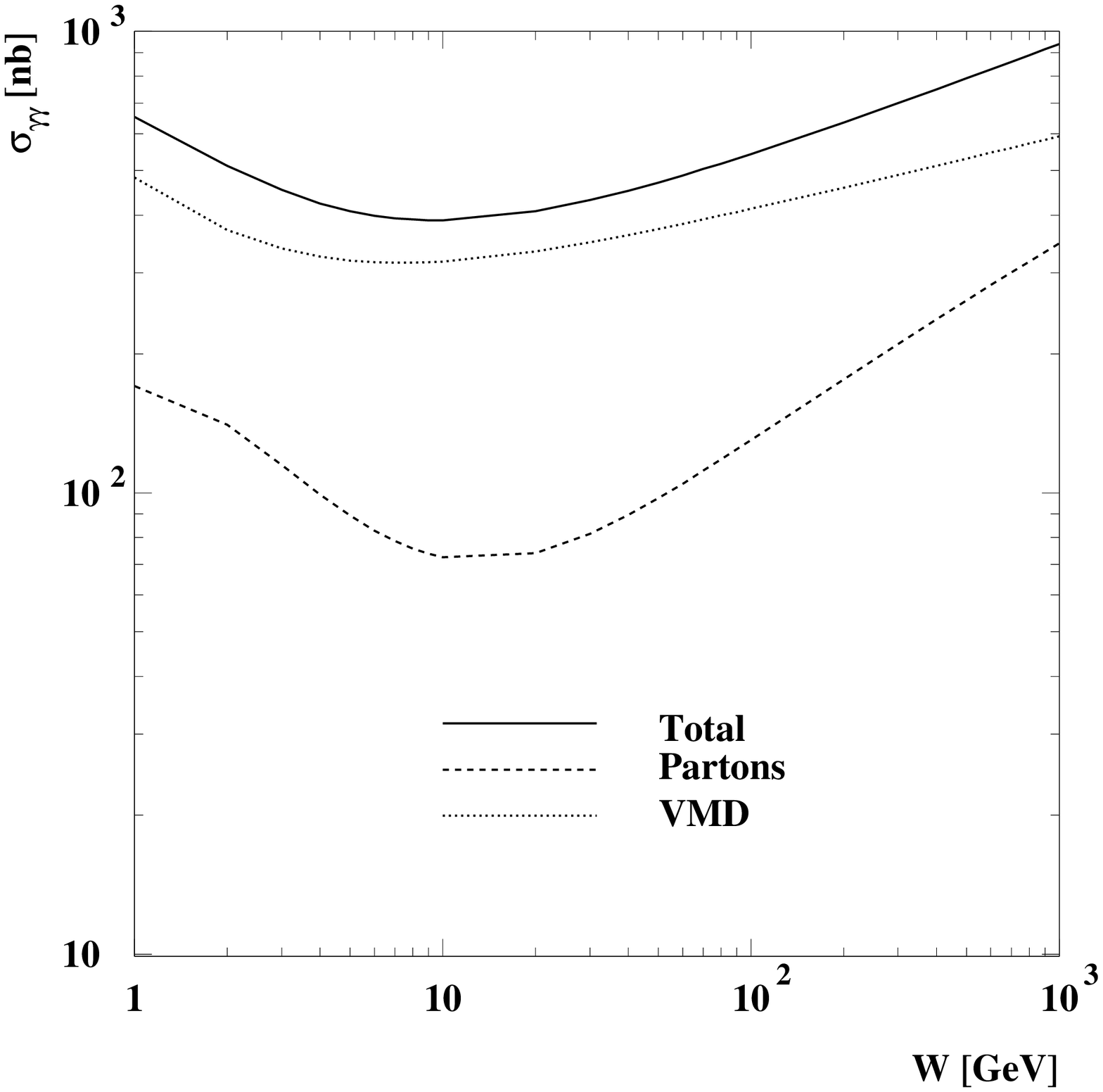,height=10cm,width=10cm}}  
\small{Figure 9: The total $\gamma \gamma$ cross-sections  
$\sigma_{\gamma \gamma}(W)$   
(continuous line)    
calculated from equation (\ref{sgamgam}) and plotted in a wide energy   
range which includes the region that will be accessible in future linear  
colliders. Shown separately are the VMD (dotted line) and partonic 
(dashed line) components of $\sigma_{\gamma \gamma}(W)$.  They correspond    
to the first and the second term of the r.h.s. of equation (\ref{sgamgam})   
respectively. Partonic contribution was obtained using the LO 
GRV \cite{GRV} parametrisation of $F_2^{\rm QCD}$}  
\label{fig:fig9}  
\end{figure}    
\noindent  
This increase is   
stronger than that implied by the soft Pomeron exchange. As a result   
the total $\gamma \gamma$ cross-section, which is the sum of the   
VMD and partonic components does also exhibit stronger increase with   
the increasing energy   
than that of the VMD component.  It is however milder than the   
increase generated by the partonic component alone,    
 at least for $W < 10^3$ GeV. This follows from the fact that    
 in this energy   
 range the magnitude of the cross-section is still dominated by its VMD   
 component.     
We found that for sufficiently high energies $W$ the total $ \gamma   
\gamma$ cross-section   
$\sigma_{\gamma \gamma}  
(W)$ described by eq. (\ref{sgamgam}) can be parametrized by  the   
effective  power   
law dependence,   $\sigma_{\gamma \gamma}  
(W) \sim (W^2)^{\lambda_{eff}}$, with $\lambda_{eff}$ slowly increasing with   
energy ($\lambda_{eff} \sim 0.1 - 0.12$ for 30 GeV $< W <$ 10$^3$   
GeV). 
\begin{figure}[htb]  
\centerline{\epsfig{file=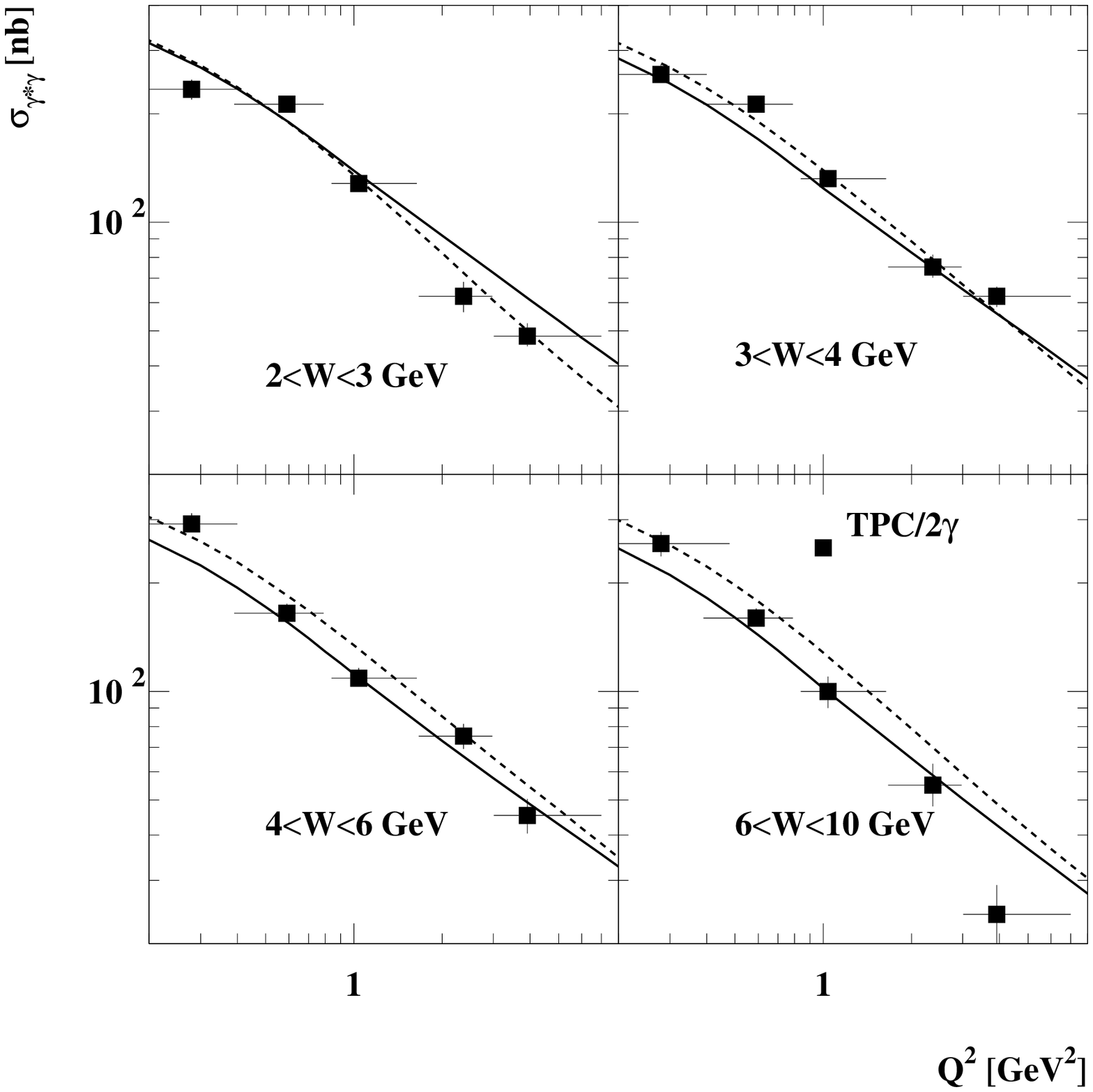,height=12cm,width=12cm}}  
\small{Figure 10: Comparison of  predictions for $\sigma_{\gamma^* \gamma}(W,Q^2)$   
in the low $Q^2$ region    
based on equations (\ref{dec}, \ref{eq:a2}, \ref{f2partons}) and (\ref{sigma})    
with  experimental results \cite{TPC}. The curves correspond to the LO 
parametrisations 
of $F_2^{\rm QCD}$: GRV \cite{GRV} (solid line) and GRS' \cite{GRS} (dashed line)  
respectively. Error bars correspond to statistical and systematic errors added 
in quadrature.  
}    
\label{fig:fig10}  
\end{figure}   
%
\begin{figure}[htb]  
\centerline{\epsfig{file=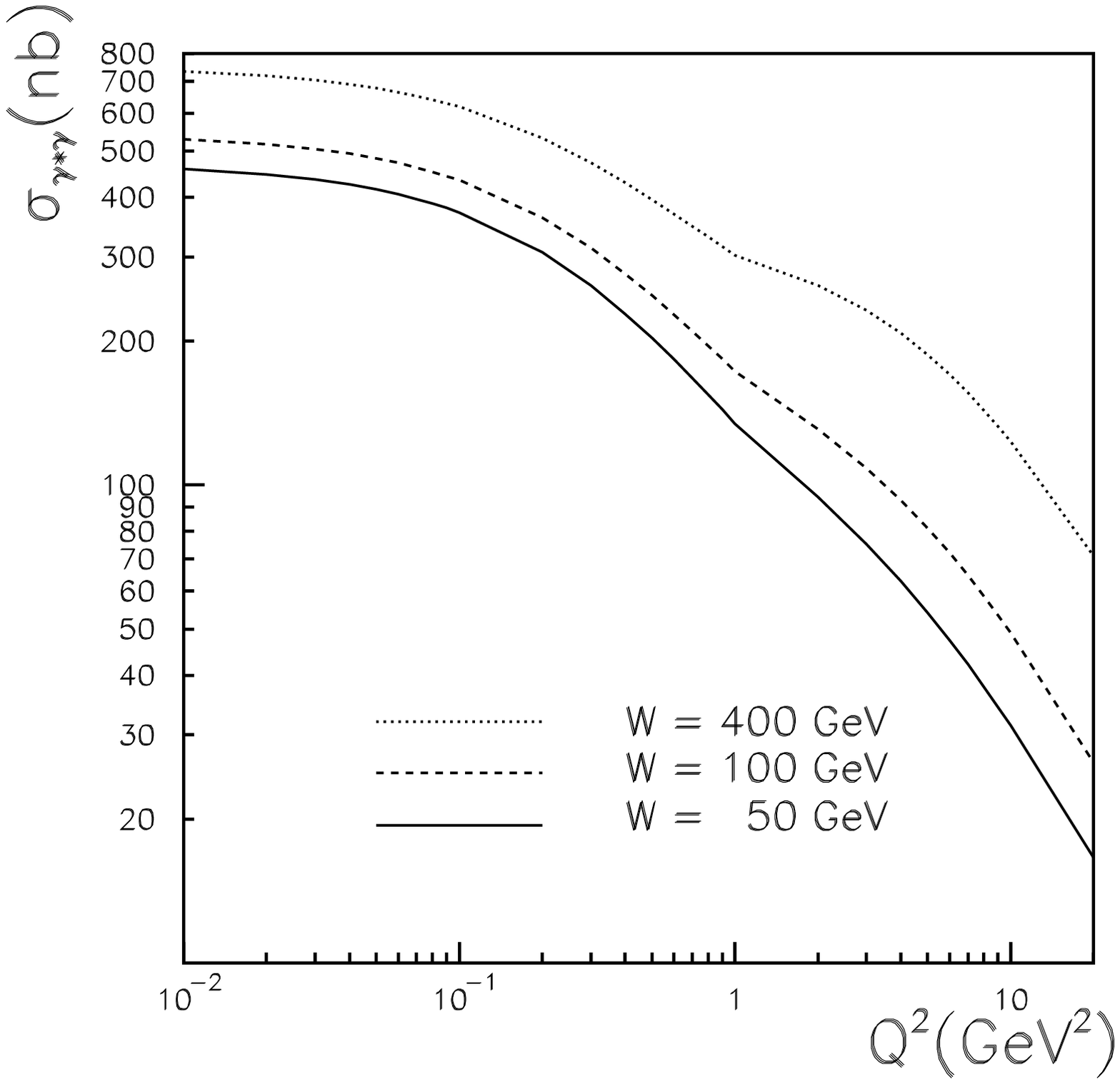,height=12cm,width=12cm}}  
\small{Figure 11: The total $\gamma^* \gamma$ cross-sections  
$\sigma_{\gamma^* \gamma}(W,Q^2)$ calculated from 
equation (\ref{sigma}) 
as a function of $Q^2$ for different values of the centre-of-mass 
 energy $W$.   
Partonic contribution was obtained using the LO GRV \cite{GRV}parametrisation of $F_2^{\rm QCD}$  
obtained from GRV \cite{GRV}.  
}    
\label{fig:fig11}  
\end{figure}  
  
In Fig.10 we show the $\gamma^*\gamma$ cross section  
for different bins of the centre-of-mass energy $W$  
plotted versus $Q^2$, the virtuality of the probing photon.  
Our theoretical predictions based on equation (\ref{sigma})   
are compared with measurements by the TPC/2$\gamma$ Collaboration  
\cite{TPC} and the agreement between the two is very good.   
In Fig.11 we show the $\gamma^*\gamma$ cross section for large  
energies $W$ plotted versus $Q^2$ (only GRV parametrsation was used here).  
At medium and large $Q^2$,   
$\sigma_{\gamma^* \gamma}$ decreases as $1/Q^2$  
(modulo logaritmic corrections)  
and for very small values ($Q^2<10^{-1} \; \rm GeV^2$) it exhibits a flattening behaviour.   
  
\section{Concluding remarks}  
\noindent  
We have presented  an extension of the representation developed for the nucleon structure function $F_2$ for arbitrary values of $Q^2$,   
\cite{BBJK1,BBJK2}, onto the structure function of the   
real photon.  This representation includes   
both the VMD contribution and the QCD component, obtained from the 
QCD parton parametrisations for the photon, 
suitably extrapolated to the region of low $Q^2$.    
In the $Q^2$=0 limit the model gives predictions for the total  
cross section $\sigma_{\gamma \gamma}$ for the interaction of two real photons.  
  
We showed that our framework   
is fairly successful in describing the experimental data   
on $\sigma_{\gamma \gamma}(W)$ and on $\sigma_{\gamma^* \gamma}(W,Q^2)$   
and on the photon structure function $F_2^{\gamma}$  
at low $Q^2$.    
We also showed that one can naturally explain the fact that the   
increase of the total $\gamma \gamma$ cross-section with increasing CM  
energy $W$  is stronger than that   
implied by soft Pomeron exchange.  The calculated total $\gamma \gamma$   
cross-section   
exhibits an approximate power-law increase with increasing energy $W$,   
i.e. $\sigma_{\gamma \gamma}(W) \sim (W^2)^{\lambda_{eff}}$   
with $\lambda_{eff}$ slowly increasing with   
energy: $\lambda_{eff} \sim 0.1 - 0.12$ for 30 GeV $< W <$ 10$^3$   
GeV.\\  
\section*{Acknowledgments}   
This research was partially supported by the Polish State Committee for   
Scientific Research grants no.    
2 P03B 089 13, 2 P03B 014 14, 2 P03B 184 10 and by the EU Fourth Framework Programme   
"Training and Mobility of Researchers",   
Network 'Quantum Chromodynamics and the Deep Structure of Elementary   
Particles', contract FMRX - CT98 - 0194.  
AMS thanks Foundation for Polish Science for financial support.   

\end{document}